\documentclass[aoas,MSNbibl,nameyear,dvips]{arximspdf}
\usepackage{algorithm,algorithmic}
\usepackage{graphicx}

\doi{10.1214/10-AOAS435}
\volume{5}
\issue{2B}
\pubyear{2011}
\firstpage{1534}
\lastpage{1552}

\makeatletter

\newcommand{\Normal}[1]{\mathcal{N}(#1)}
\newcommand{\GDist}[1]{\operatorname{Gamma}(#1)}
\renewcommand{\v}[1]{\mathbf{#1}}
\newcommand{\expb}[1]{\exp{(#1)}}
\newcommand{\G}{\v{G}}
\newcommand{\Y}{\v{Y}}
\newcommand{\Z}{\v{Z}}
\newcommand{\X}{\v{X}}
\newcommand{\xt}{\v{x}_n}
\newcommand{\et}{\bolds{\epsilon}_n}
\newcommand{\zk}{\v{z}_{-kn}}

\let\epsilon\varepsilon

\makeatother

\begin{document}
\begin{frontmatter}

\title{Nonparametric Bayesian sparse factor models with application to
gene expression modeling}
\runtitle{Nonparametric Bayesian sparse factor models}

\begin{aug}
\author[A]{\fnms{David} \snm{Knowles}\corref{}\thanksref{t1}\ead[label=e1]{dak33@cam.ac.uk}}
and
\author[A]{\fnms{Zoubin} \snm{Ghahramani}\thanksref{t2}\ead[label=e2]{zoubin@eng.cam.ac.uk}}

\thankstext{t1}{Supported by Microsoft Research through the Roger
Needham Scholarship at Wolfson College, University of Cambridge. }
\thankstext{t2}{Supported by EPSRC Grant EP/F027400/1.}

\runauthor{D. Knowles and Z. Ghahramani}

\affiliation{University of Cambridge}

\address[A]{Engineering Department \\
Cambridge University\\
Trumpington Street\\
Cambridge, CB2 1PZ\\
United Kingdom\\
\printead{e1}\\
\phantom{E-mail:\ }\printead*{e2}} %adresu isvedimo komanda gale!
\end{aug}

% HISTORY:
\received{\smonth{6} \syear{2010}}
\revised{\smonth{10} \syear{2010}}

% ABSTRACT
%
\begin{abstract}
A nonparametric Bayesian extension of Factor Analysis (FA) is proposed
where observed data $\mathbf{Y}$ is modeled as a linear superposition,
$\mathbf{G}$, of
a potentially infinite number of hidden factors, $\mathbf{X}$. The Indian
Buffet Process (IBP) is used as a prior on $\mathbf{G}$ to incorporate sparsity
and to allow the number of latent features to be inferred. The model's
utility for modeling gene expression data is investigated using
randomly generated data sets based on a known sparse connectivity
matrix for \textit{E. Coli}, and on three biological data sets of
increasing complexity.
\end{abstract}

% KEYWORDS
%

\end{frontmatter}

%s1 ###
\section{Introduction}

Principal Components Analysis (PCA), Factor Analysis (FA) and
Independent Components Analysis (ICA) are models which explain observed
data, $\v{y}_n\in\mathbb{R}^D$, in terms of a linear superposition
of independent hidden factors, $\xt\in\mathbb{R}^K$, so
%
%e1 ###
\begin{equation}  \label{eq:fa}
\v{y}_n= \G\xt+ \et,
\end{equation}
where $\G$ is the factor loading matrix and $\et$ is a noise vector,
usually assumed to be Gaussian. These algorithms can be expressed in
terms of performing inference in appropriate probabilistic models. The
latent factors are usually considered as random variables, and the
mixing matrix as a parameter to estimate. In both PCA and FA the latent
factors are given a standard (zero mean, unit variance) normal prior.
In PCA the noise is assumed to be isotropic, whereas in FA the noise
covariance is only constrained to be diagonal. A standard approach in
these models is to integrate out the latent factors and find the
maximum likelihood estimate of the mixing matrix. In ICA the latent
factors are assumed to be heavy-tailed, so it is not usually possible
to integrate them out. In this paper we take a fully Bayesian approach,
viewing not only the hidden factors but also the mixing coefficients as
random variables whose posterior distribution given data we aim to infer.

Sparsity plays an important role in latent feature models, and is
desirable for several reasons. It gives improved predictive
performance, because factors irrelevant to a particular dimension are
not included. Sparse models are more readily interpretable since a
smaller number of factors are associated with observed dimensions. In
many real world situations there is an intuitive reason why we expect
sparsity: for example, in gene regulatory networks a transcription
factor will only regulate genes with specific motifs. In our previous
work [\citet{knowles07iica}] we investigated the use of sparsity on the
latent factors $\xt$, but this formulation is not appropriate in the
case of modeling gene expression, where, as described above, a
transcription factor will regulate only a small set of genes,
corresponding to sparsity in the factor loadings, $\G$. Here we
propose a novel approach to sparse latent factor modeling where we
place sparse priors on the factor loading matrix, $\G$. The Bayesian
Factor Regression Model of \citet{westBFRM2008} is closely related to
our work in this way, although the hierarchical sparsity prior they use
is somewhat different. An alternative ``soft'' approach to
incorporating sparsity is to put a $\operatorname{Gamma}(a,b)$ (usually
exponential, i.e., $a=1$) prior on the precision of each element of $\G
$ independently, resulting in the elements of $\G$ being marginally
Student-$t$ distributed a priori; see \citet{fokoue}, \citet{godsill},
and \citet{arch08sparseFA}. A LASSO-based approach to generating a
sparse factor loading has also been developed [\citet{TibshiraniSPCA2006}; \citet{Hastie2009}]. We compare these sparsity schemes
empirically in the context of gene expression modeling.

A problematic issue with this type of model is how to choose the latent
dimensionality of the factor space, $K$. Model selection can be used to
choose between different values of $K$, but generally requires
significant manual input and still requires the range of $K$ over which
to search to be specified. \citet{ZhangRjmcmcPca} applied Reversible
Jump MCMC to PCA, which has many of the advantages of our approach: a
posterior distribution over the number of latent dimensions can be
approximated, and the number of latent dimensions could potentially be
unbounded. However, RJ MCMC is considerably more complex to implement
for sparse Factor Analysis than our proposed framework.

We use the Indian Buffet Process [\citet{ibp}], which defines a
distribution over infinite binary matrices, to provide sparsity and a
framework for inferring the appropriate latent dimension of the data
set using a straightforward Gibbs sampling algorithm. The Indian Buffet
Process (IBP) allows a \mbox{potentially} unbounded number of latent factors,
so we do not have to specify a maximum number of latent dimensions a
priori. We denote our model ``NSFA'' for ``Nonparametric Sparse Factor
Analysis.'' Our model is closely related to that of \citet{daume08ihfrm}, and is a~simultaneous development.

%s2 ###
\section{The model}

We will define our model in terms of equation (\ref{eq:fa}). Let~$\Z$
be a binary matrix whose $(d,k)$th element represents whether observed
dimension $d$ includes any contribution from factor $k$. We then model
the mixing matrix by
%
%e2 ###
\begin{equation} \label{eq:sparsity}
p(g_{dk}|Z_{dk},\lambda_k)=Z_{dk} \Normal{g_{dk};0,\lambda
_k^{-1}}+(1-Z_{dk})\delta_0(g_{dk}),
\end{equation}
where $\lambda_k$ is the inverse variance (precision) of the $k$th
factor and $\delta_0$ is a delta function (point-mass) at 0.
Distributions of this type are sometimes known as ``spike and slab''
distributions. We allow a potentially infinite number of hidden
sources, so that $\Z$ has infinitely many columns, although only a~%
finite number will have nonzero entries. This construction allows us to
use the IBP to provide sparsity and define a generative process for the
number of latent factors.

\begin{figure}[b]

\includegraphics{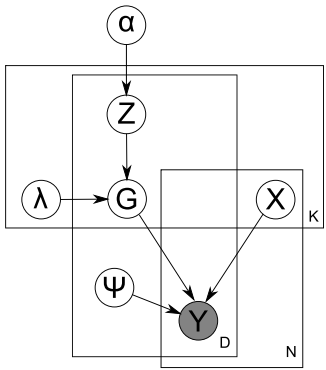}

\caption{Graphical model.}
\label{fig:iICA}
\end{figure}

We assume independent Gaussian noise, $\et$, with diagonal covariance
matrix $\Psi$. We find that for many applications assuming isotropic
noise is too restrictive, but this option is available for situations
where there is strong prior belief that all observed dimensions should
have the same noise variance. The latent factors, $\xt$, are given
Gaussian priors. Figure \ref{fig:iICA} shows the complete graphical model.

%f1 ###

%s2.1 ###
\subsection{Defining a distribution over infinite binary matrices}\label{sec:ibp}
We now define our infinite model by taking the limit of a series of
finite models.

\subsubsection*{Start with a finite model}

We derive the distribution on $\Z$ by defining a finite $K$ model and
taking the limit as $K\rightarrow\infty$. We then show how the
infinite case corresponds to a simple stochastic process.

We have $D$ dimensions and $K$ hidden sources. Recall that $z_{dk}$ of
matrix~$\Z$ tells us whether hidden source $k$ contributes to
dimension $d$. We assume that the probability of a source $k$
contributing to any dimension is $\pi_k$, and that the rows are
generated independently. We find
%
%e3 ###
\begin{equation}
P(\Z|\bolds{\pi})=\prod_{k=1}^K\prod_{d=1}^D P(z_{dk}|\pi_k)=\prod
_{k=1}^K\pi_k^{m_k}(1-\pi_k)^{D-m_k},
\end{equation}
where $m_k=\sum_{d=1}^D z_{dk}$ is the number of dimensions to which
source $k$ contributes. The inner term of the product is a binomial
distribution, so we choose the conjugate Beta$(r,s)$ distribution for
$\pi_k$. For now we take $r=\frac{\alpha}{K}$ and $s=1$, where
$\alpha$ is the strength parameter of the IBP. The model is defined by
%
%e5 ###
%e4 ###
\begin{eqnarray}  \label{eq:ibpsimple}
\pi_k | \alpha&\sim&\operatorname{Beta} \biggl(\frac{\alpha}{K},1 \biggr), \\
z_{dk} | \pi_k &\sim&\operatorname{Bernoulli}(\pi_k).
\end{eqnarray}
Due to the conjugacy between the binomial and beta distributions, we
are able to integrate out $\pi$ to find
%
%e6 ###
\begin{equation} \label{eq:finite}
P(\Z)=\prod_{k=1}^K\frac{(\alpha/K)\Gamma(m_k+
{\alpha}/{K})\Gamma(D-m_k+1)}{\Gamma(D+1+{\alpha}/{K})},
\end{equation}
where $\Gamma(\cdot)$ is the Gamma function.

\subsubsection*{Take the infinite limit}

Griffiths and Ghahramani (\citeyear{ibp}) define a scheme to order the nonzero rows of $\Z$ which
allows us to take the limit $K\rightarrow\infty$ and find
%
%e7 ###
\begin{equation} \label{eq:ibpfull}
P(\Z)=\frac{\alpha^{K_+}}{\prod_{h>0}K_h!}\expb{-\alpha H_D}\prod
_{k=1}^{K_+}\frac{(D-m_k)!(m_k-1)!}{N!},
\end{equation}
where $K_+$ is the number of active features (i.e., nonzero columns of
$\Z$), $H_D=\sum_{j=1}^D\frac{1}{j}$ is the $D$th harmonic number,
and $K_h$ is the number of rows whose entries correspond to the binary
number $h$.

\subsubsection*{Go to an Indian Buffet}

\begin{figure}
\begin{tabular}{@{}cc@{}}

\includegraphics{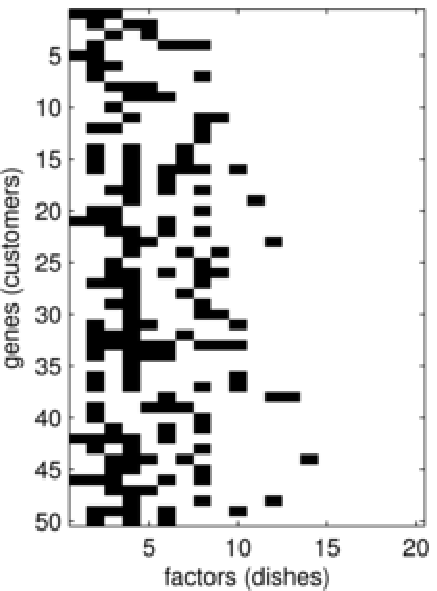}
&\includegraphics{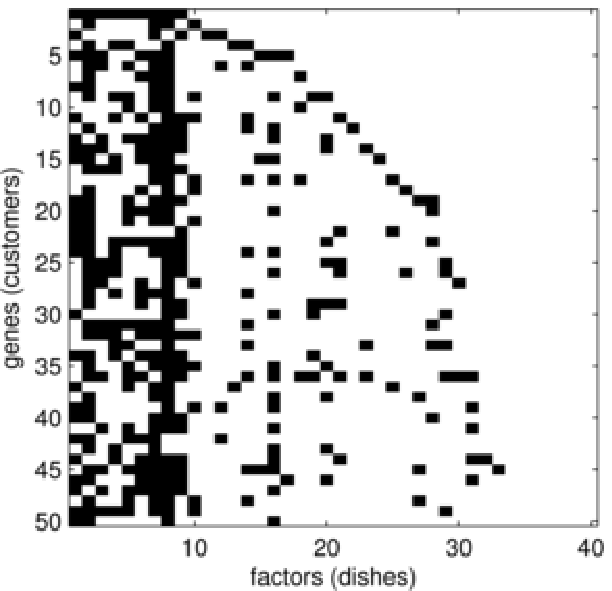}\\
(a)&(b)
\end{tabular}
\caption{Draws from the one parameter IBP for two different values of
$\alpha$. \textup{(a)} $\alpha=4$. \textup{(b)}~$\alpha=8$.} \label{fig:ibp1}
\end{figure}

This distribution corresponds to a simple stochastic process, the
Indian Buffet Process. Consider a buffet with a seemingly infinite
number of dishes (hidden sources) arranged in a line. The first
customer (observed dimension) starts at the left and samples
$\operatorname{Poisson}(\alpha)$ dishes. The $i$th customer
moves from left to right sampling dishes with probability $\frac
{m_k}{i}$ where $m_k$ is the number of customers to have previously
sampled dish $k$. Having reached the end of the previously sampled
dishes, he tries $\operatorname{Poisson}(\frac{\alpha}{i})$ new dishes.
Figure \ref{fig:ibp1} shows two draws from the IBP for two different
values of $\alpha$.

%f2 ###

If we apply the same ordering scheme to the matrix generated by this
process as for the finite model, we recover the correct exchangeable
distribution. Since the distribution is exchangeable with respect to
the customers, we find by considering the last customer that
%
%e8 ###
\begin{equation}
P(z_{kt}=1|\zk) = \frac{m_{k,-t}}{D}, \label{eq:ibp}
\end{equation}
where $m_{k,-t} = \sum_{s \ne t} z_{ks}$, which is used in sampling
$\Z$. By exchangeability and considering the first customer, the
number of active sources for each dimension follows a Poisson$(\alpha
)$ distribution, and the expected number of entries in $\Z$ is
$D\alpha$. We also see that the number of active features $K_+=\sum
^D_{d=1}\operatorname{Poisson}(\frac{\alpha}{d})=\operatorname{Poisson}(\alpha H_D)$.

%s3 ###
\section{Related work}

The Bayesian Factor Regression Model (BFRM) of \citet{westBFRM2008} is
closely related to the finite version of our model. The key difference
is the use of a hierarchical sparsity prior. Each element of $\G$ has
prior of the form
\[
g_{dk} \sim(1-\pi_{dk})\delta_0(g_{dk}) + \pi_{dk} \Normal
{g_{dk};0,\lambda_k^{-1}}.
\]
The finite IBP model is equivalent to setting $ \pi_{dk} = \pi_k \sim
\operatorname{Beta}(\alpha/K,1) $ and then integrating out $\pi_k$. In BFRM
a hierarchical prior is used:
\[
\pi_{dk} \sim(1-\rho_k) \delta_0 (\pi_{dk}) + \rho_k \operatorname{Beta}\bigl(\pi_{dk}; am,
a(1-m)\bigr),
\]
where $\rho_k \sim\operatorname{Beta}(sr,s(1-r))$. Nonzero elements of $\pi
_{dk}$ are given a diffuse prior favoring larger probabilities
[$a=10$, $m=0.75$ are suggested in \citet{westBFRM2008}], and $\rho_k$ is
given a prior which strongly favors small values, corresponding to a
sparse solution (e.g., $s=D$, $r=\frac{5}{D}$).

Note that on integrating out $\pi_{dk}$, the prior on $g_{dk}$ is
\[
g_{dk} \sim(1-m\rho_k)\delta_0(g_{dk}) + m\rho_k \Normal
{g_{dk};0,\lambda_k^{-1}}.
\]

This hierarchical sparsity prior is motivated by improved
interpretability in terms of less uncertainty in the posterior as to
whether an element of $G$ is nonzero. However, this comes at a cost of
significantly increased computation and reduced predictive performance,
suggesting that the uncertainty removed from the posterior was actually
important.

The LASSO-based Sparse PCA (SPCA) method of \citet{TibshiraniSPCA2006}
and \citet{Hastie2009} has similar aims to our work in terms of
providing a sparse variant of PCA to aid interpretation of the results.
However, since SPCA is not formulated as a generative model, it is not
necessarily clear how to choose the regularization parameters or
dimensionality without resorting to cross-validation. In our
experimental comparison to SPCA, we adjust the regularization constants
such that each component explains roughly the same proportion of the
total variance as the corresponding standard (nonsparse) principal component.

%s4 ###
\section{Inference} \label{sec:inference}

Given the observed data $\Y$, we wish to infer the hidden sources $\X
$, which sources are active $\Z$, the mixing matrix $\G$, and all
hyperparameters. We use Gibbs sampling, but with Metropolis--Hastings
(MH) steps for sampling new features. We draw samples from the marginal
distribution of the model parameters given the data by successively
sampling the conditional distributions of each parameter in turn, given
all other parameters.\looseness=-1

Since we assume independent Gaussian noise, the likelihood function is
\begin{eqnarray}  \label{eq:likelihood}
P(\Y|\G,\X,\bolds{\psi}) &=& \prod_{n=1}^N \frac{1}{(2 \pi)^
{D/2} |\bolds{\psi}|^{1/2}}\nonumber\\[-8pt]\\[-8pt]
&&{}\times\exp\biggl(-\frac{1}{2} (\v{y}_n- \G\xt)^T\bolds{\psi}^{-1} (\v{y}_n-
\G\xt)\biggr),\nonumber
\end{eqnarray}
where $\bolds{\psi}$ is a diagonal noise covariance matrix.

\subsection*{Notation}

We use $-$ to denote the ``rest'' of the model,
that is, the values of all variables not explicitly conditioned upon in
the current state of the Markov chain. The $r$th row and $c$th column
of matrix $A$ are denoted $A_{r:}$ and $A_{:c}$ respectively.

\subsection*{Mixture coefficients}

We first derive a Gibbs sampling step
for an individual element of the IBP matrix, $Z_{dk}$, determining
whether factor $k$ is active for dimension $d$. Recall that $\lambda
_k$ is the precision (inverse covariance) of the factor loadings for
the $k$th factor. The ratio of likelihoods can be calculated using
equation (\ref{eq:likelihood}). Integrating out the $(d,k)$th element of
the factor loading matrix $g_{dk}$ [whose prior is given by equation
(\ref{eq:sparsity})], we obtain
%
%e10 ###
%e9 ###
\begin{eqnarray}
\frac{P(\Y|Z_{dk}=1,-)}{P(\Y|Z_{dk}=0,-)} &=& \frac{\int P(\Y
|g_{dk},-) \Normal{g_{dk};0,\lambda^{-1}_k}\, dg_{dk}}{P(\Y
|g_{dk}=0,-)}\\
\label{eq:rl}
&=& \sqrt{\frac{\lambda_k}{\lambda}} \exp\biggl(\frac12 \lambda\mu
^2\biggr),
\end{eqnarray}
where we have defined $\lambda=\psi^{-1}_d X_{k:}^T X_{k:}+\lambda
_k$ and $\mu=\frac{\psi^{-1}_d}{\lambda} X_{k:}^T \hat{E}_{d:}$
with the matrix of residuals $\hat{\v{E}}=\Y-\G\X$ evaluated with
$g_{dk}=0$. The dominant calculation is that for $\mu$ since the
calculation for $\lambda$ can be cached. This operation is $O(N)$ and
must be calculated $D\times K$ times, so sampling the IBP matrix, $\Z
$, and factor loading matrix, $\G$, is order $O(NDK)$.

From the exchangeability of the IBP, we can imagine that dimension $d$
was the last to be observed, so that the ratio of the priors is
%
%e11 ###
\begin{equation} \label{eq:rp}
\frac{P(Z_{dk}=1|-)}{P(Z_{dk}=0|-)}=\frac{m_{-d,k}}{N-1-m_{-d,k}},
\end{equation}
where $m_{-d,k}$ is the number of dimensions for which factor $k$ is
active, excluding the current dimension $d$. Multiplying equations (\ref
{eq:rl}) and (\ref{eq:rp}) gives the expression for the ratio of
posterior probabilities for $Z_{dk}$ being 1 or 0, which is used for
sampling. If $Z_{dk}$ is set to 1, we sample $g_{dk}|- \sim\Normal
{\mu,\lambda^{-1}}$ with $\mu,\lambda$ defined as for equation~(\ref{eq:rl}).

\subsection*{Adding new features}

\begin{figure}

\includegraphics{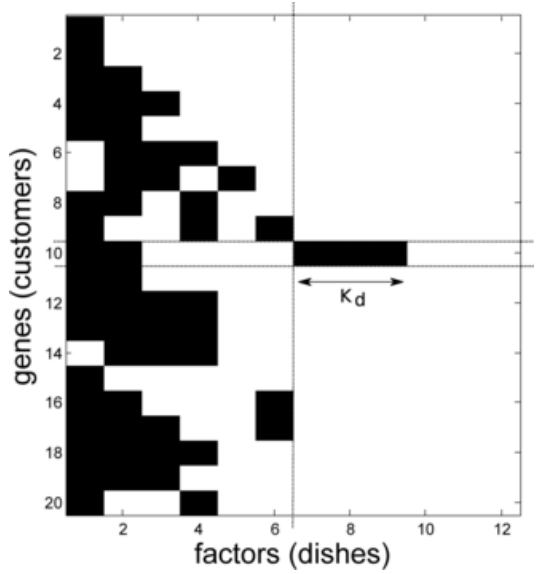}

\caption{A diagram to illustrate the definition of $\kappa_d$, for $d=10$.}
\label{fig:sampleK}
\end{figure}

$\Z$ is a matrix with infinitely many
columns, but the nonzero columns contribute to the likelihood and are
held in memory. However, the zero columns still need to be taken into
account since the number of active factors can change. Let $\kappa_d$
be the number of columns of~$\Z$ which contain 1 only in row $d$, that
is, the number of features which are active only for dimension $d$.
Note that due to the form of the prior for elements of $\Z$ given in
equation (\ref{eq:rp}), $\kappa_d=0$ for all $d$ after a sampling sweep
of $\Z$. Figure \ref{fig:sampleK} illustrates $\kappa_d$ for a
sample $\Z$ matrix.

%f3 ###

New features are proposed by sampling $\kappa_d$ with a MH step. It is
possible to integrate out either the new elements of the mixing matrix,
$\v{g}$ (a $1\times\kappa_d$ vector), or the new rows of the latent
feature matrix, $\X'$ (a $\kappa_d\times N$ matrix), but not both.
Since the latter generally has higher dimension, we choose to integrate
out $\X'$ and include $\v{g}^T$ as part of the proposal. Thus, the
proposal is $\xi=\{\kappa_d,\v{g}\}$, and we propose a move $\xi
\rightarrow\xi^*$ with probability $J(\xi^*|\xi)$. In this case
$\xi=\varnothing$ since, as noted above, $\kappa_d=0$ initially. The
simplest proposal, following \citet{meeds}, would be to use the prior
on $\xi^*$, that is,
\[
J(\xi)=P(\kappa_d|\alpha)\cdot p(\v{g}|\kappa_d,\lambda_k)=\operatorname{Poisson} (\kappa_d;\gamma) \cdot
N(\v{g};0,\lambda_k^{-1}),
\]
where $\gamma=\frac{\alpha}{D-1}$.

Unfortunately, the rate constant of the Poisson prior tends to be so
small that new features are very rarely proposed, resulting in slow
mixing. To remedy this, we modify the proposal distribution for $\kappa
_d$ and introduce two tunable parameters, $\pi$ and $\lambda$:
%
%e12 ###
\begin{equation} \label{eq:kproposal}
J(\kappa_d) = (1-\pi)\operatorname{Poisson} (\kappa_d;\lambda\gamma)+\pi\mathbf{1}(\kappa_d=1).
\end{equation}

Thus, the Poisson rate is multiplied by a factor $\lambda$, and a
spike at $\kappa_d=1$ is added with mass $\pi$. The proposal is
accepted with probability $\min{(1,a_{\xi\rightarrow\xi^*})}$ where
\begin{eqnarray}  \label{eq:r}
a_{\xi\rightarrow\xi^*}&=&
\frac{P(\xi^*|-,Y)J(\xi|\xi^*)}
{P(\xi|-,Y)J(\xi^*|\xi)}\nonumber\\[-8pt]\\[-8pt]
&=&\frac{P(Y|\xi^*,-)P(\kappa_d|\alpha)p(\v{g}|\kappa_d,\lambda
_k)}{P(Y|-)J(\kappa_d)p(\v{g}|\kappa_d,\lambda_k)}=a_l \cdot a_p,\nonumber
\end{eqnarray}
where
%
%e14 ###
%e13 ###
\begin{eqnarray}
a_l &=& \frac{P(Y|\xi^*,-)}{P(Y|-)},\\
a_p &=& \frac{P(\kappa_d|\alpha)}{J(\kappa_d)} = \frac{\operatorname{Poisson} (\kappa_d;\gamma)}{\operatorname{Poisson} (\kappa_d;\lambda
\gamma)}.
\end{eqnarray}
Note that we take $J(\xi|\xi^*)=1$ since $\xi=\varnothing$. To
calculate the likelihood ratio,~$a_l$, we need the collapsed likelihood
under the new proposal:
%
%e15 ###
\begin{eqnarray}
P(Y_{d:}|\xi^*,-)&=&\prod_{n=1}^N \int P(Y_{dn}|\xi^*,\v
{x}'_n,-)P(\v{x}'_n)\,d\v{x}'\\
&=& \prod_{n=1}^N (2\pi\psi_d^{-1})^{-{1/2}} (2\pi)^
{\kappa_d/2} |\v{M}|^{-{1/2}} \nonumber\\[-8pt]\\[-8pt]
&&{}\times\exp\biggl(\frac{1}{2}(\v
{m}_n^T\v{M}\v{m}_n-\psi_d^{-1} \hat{E}^2_{dn})\biggr),\nonumber
\end{eqnarray}
where we have defined $\v{M}=\psi_d^{-1}\v{g}\v{g}^T+I_{\kappa_d}$
and $\v{m}_n=\v{M}^{-1}\psi_d^{-1}\v{g}\hat{E}_{dn}$ with the
matrix of residuals $\hat{\v{E}}=\Y-\G\X$. The likelihood under
the current sample~is
%
%e16 ###
\begin{equation}
P(Y_{d:}|\xi,-)= \prod_{n=1}^N (2\pi\psi_d^{-1})^{-{1/2}}
\exp\biggl(-\frac{1}{2}\psi_d^{-1} \hat{E}^2_{dn}\biggr).
\end{equation}
Substituting these likelihood terms into the expression for the ratio
of likelihood terms, $a_l$, gives
%
%e17 ###
\begin{equation}
a_l= (2\pi)^{N\kappa_d/2} |\v{M}|^{-{N/2}} \exp\biggl(
\frac{1}{2}\sum_n\v{m}_n^T\v{M}\v{m}_n\biggr).
\end{equation}

We found that appropriate scheduling of the sampler improved mixing,
particularly with respect to adding new features. The final scheme we
settled on is described in Algorithm \ref{alg:sample}.

\begin{algorithm}
\caption{One iteration of the NSFA sampler}
\label{alg:sample}
\begin{algorithmic}
\FOR{$d=1$ to $D$}
\FOR{$k=1$ to $K$}
\STATE Sample $Z_{dk}$
\ENDFOR
\STATE Sample $\kappa_d$
\ENDFOR
\FOR{$n=1$ to $N$}
\STATE Sample $X_{:n}$
\ENDFOR
\STATE Sample $\alpha,\phi,\lambda_g$
\end{algorithmic}
\end{algorithm}

\subsection*{IBP parameters}

We can choose to sample the IBP strength
parameter $\alpha$, with conjugate $\operatorname{Gamma}(e,f)$ prior
(note that we use the inverse scale parameterization of the Gamma
distribution). The conditional prior of equation (\ref{eq:ibpfull}) acts
as the likelihood term and the posterior update is as follows:
%
%e18 ###
\begin{equation}
P(\alpha|\Z) \propto P(\Z|\alpha) P(\alpha) = \GDist{\alpha;K_+
+e,f+H_D},
\label{eq:samplealpha}
\end{equation}
where $K_+$ is the number of active sources and $H_D=\sum_{j=1}^D\frac
{1}{j}$ is the $D$th harmonic number.

The remaining sampling steps are standard, but are included here for
completeness.

\subsection*{Latent variables}

Sampling the columns $\v{x}_n$ of the
latent variable matrix~$\X$ for each $t \in[1,\dots,N]$, we have
%
%e19 ###
\begin{equation}
P(\v{x}_n|-)\propto P(\v{y}_n|\v{x}_n,-)P(\v{x}_n)=\Normal{\v
{x}_n;\bolds{\mu}_n,\bolds{\Lambda}},
\end{equation}
where we have defined $\bolds{\Lambda}=\G^T\bolds{\psi}^{-1}\G+I$ and
$\bolds{\mu}_n=\bolds{\Lambda}^{-1}\G^T\bolds{\psi}^{-1}\v{y}_n$. Note
that since $\bolds{\Lambda}$ does not depend on $n$, we only need to
compute and invert it once per iteration. Calculating $\bolds{\Lambda}$
is order $O(K^2 D)$, and inverting it is~$O(K^3)$. Calculating $\bolds{\mu}_t$ is order $O(KD)$ and must be calculated for all $N$ $\v
{x}_t$'s, a total of $O(NKD)$. Thus, sampling $\X$ is order $O(K^2+K^3+NKD)$.

\subsection*{Factor precision}

If the mixture coefficient prior
precisions $\lambda_k$ are constrained to be equal, we have $\lambda
_k = \lambda\sim\operatorname{Gamma}(c,d)$. The posterior update is
then given by $\lambda|G \sim\operatorname{Gamma}(c+ \frac{\sum_k
m_k}{2},d+\sum_{d,k} G^2_{dk})$.

However, if the variances are allowed to be different for each column
of~$\G$, we set $\lambda_k \sim\operatorname{Gamma}(c,d)$, and the
posterior update is given by $\lambda_k|G \sim\operatorname
{Gamma}(c+\frac{m_k}{2},d+\sum_d G^2_{dk})$. In this case we may also
wish to share power across factors, in which case we also sample $d$.
Putting a Gamma prior on~$d$ such that $d \sim\operatorname
{Gamma}(c_0,d_0)$, the posterior update is $d|\lambda_k \sim
\operatorname{Gamma}(c_0+cK,d_0+\sum_{k=1}^K\lambda_k)$.

\subsection*{Noise variance}

The additive Gaussian noise can be
constrained to be isotropic, in which case the inverse variance is
given a Gamma prior: $\psi_d^{-1}=\psi^{-1}\sim\operatorname
{Gamma}(a,b)$ which gives the posterior update $\psi^{-1}|-\sim
\operatorname{Gamma}(a+\frac{ND}{2},b+\sum_{d,n}\hat{E}^2_{dn})$.

However, if the noise is only assumed to be independent (which we have
found to be more appropriate for gene expression data), then each
dimension has a separate noise variance, whose inverse is given a Gamma
prior: $\psi_d^{-1}\sim\operatorname{Gamma}(a,b)$ which gives the
posterior update $\psi_d^{-1}|-\sim\operatorname{Gamma}(a+\frac
{N}{2},b+\sum_n E^2_{dn})$ where the matrix of residuals $\hat{\v
{E}}=\Y-\G\X$. We can share power between dimensions by giving the
hyperparameter $b$ a hyperprior $\operatorname{Gamma}(a_0,b_0)$
resulting in the Gibbs update $b|-\sim\operatorname
{Gamma}(a_0+aD,b_0+\sum_{d=1}^D \psi_d^{-1})$. This hierarchical
prior results in soft coupling between the noise variances in each
dimension, so we will refer to this variant as \textit{sc}.

%s5 ###
\section{Results}

We compare the following models:
\begin{itemize}
\item FA---Bayesian Factor Analysis; see, for example, \citet{bfa} or
\citet{bfa2}.
\item AFA---Factor Analysis with ARD prior to determine active
sources.
\item FOK---The sparse Factor Analysis method of \citet{fokoue}, \citet
{godsill} and \citet{arch08sparseFA}.
\item SPCA---The Sparse PCA method of
\citet{TibshiraniSPCA2006}.
\item BFRM---Bayesian Factor Regression Model of
\citet{westBFRM2008}.
\item SFA---Sparse Factor Analysis, using the finite IBP.
\item NSFA---The proposed model: Nonparametric Sparse Factor Analysis.
\end{itemize}

Note that all of these models can be learned using the software package
we provide simply by using appropriate settings.

%s5.1 ###
\subsection{Synthetic data} \label{sec:syndata}

Since generating a connectivity matrix $\Z$ from the IBP itself would
clearly bias toward our model, we instead use the $D=100$ gene by
$K=16$ factor \emph{E. Coli} connectivity matrix derived in \citet
{Kao2004} from RegulonDB and current literature. We ignore whether the
connection is believed to be up or down regulation, resulting in a
binary matrix $\Z$. We generate random data sets with $N=100$ samples
by drawing the nonzero elements of $\G$ (corresponding to the elements
of $\Z$ which are nonzero), and all elements of $\X$, from a zero
mean unit variance Gaussian, calculating ${\Y=\G\X+\v{E}}$, where
$\v{E}$ is Gaussian white noise with variance set to give a signal to
noise ratio of 10.

Here we will define the reconstruction error $E_r$ as
\[
E_r(G,\hat{G})=\frac{1}{DK}\sum_{k=1}^K \min_{\hat{k}\in\{
1,\ldots,\hat{K}\}} \sum_{d=1}^D(G_{dk}-G_{d\hat{k}})^2,
\]
where $\hat{G},\hat{K}$ are the inferred quantities. Although we
minimize over permutations, we do not minimize over rotations since, as
noted in \citet{fokoue}, the sparsity of the prior stops the solution
being rotation invariant. We average this error over the last ten
samples of the MCMC run. This error function does not penalize
inferring extra spurious factors, so we will investigate this
possibility separately. The precision and recall of active elements of
the $\Z$ achieved by each algorithm (after thresholding for the
nonsparse algorithms) are presented in the Supplementary Material, but omitted here since the results are consistent with
the reconstruction error.

%, so that the model is identifiable up to permutation of the factors

%f4 ###
\begin{figure}

\includegraphics{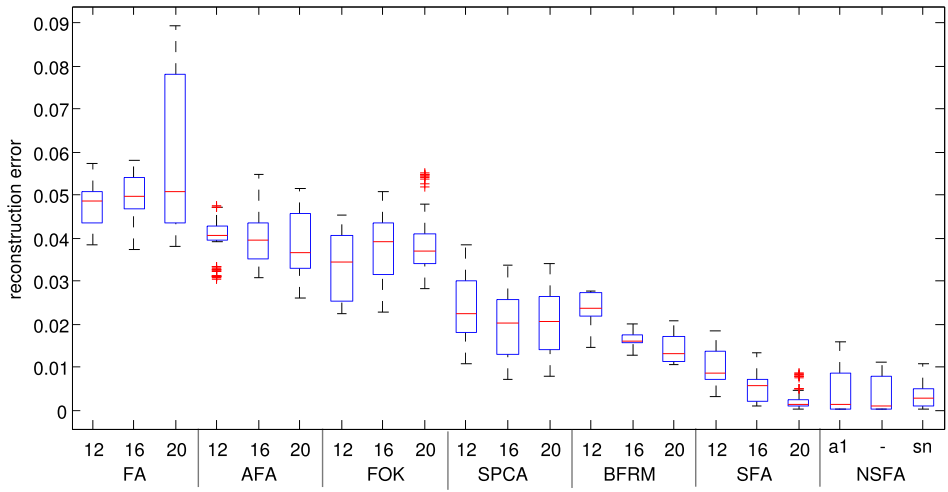}

\caption{Boxplot of reconstruction errors for simulated data derived
from the \emph{E. Coli} connectivity matrix of Kao et~al. (\protect\citeyear{Kao2004}). Ten
data sets were generated and the reconstruction error calculated for
the last ten samples from each algorithm. Numbers refer to the number
of latent factors used, $K$. \emph{a1} denotes fixing $\alpha=1$.
\emph{sn} denotes sharing power between noise dimensions.}
\label{fig:syn_test}
\end{figure}

The reconstruction error for each method with different numbers of
latent features is shown in Figure \ref{fig:syn_test}. Ten random data
sets were used and for the sampling methods (all but SPCA) the results
were averaged over the last ten samples out of 1000. Unsurprisingly,
plain Factor Analysis (FA) performs the worst, with increasing
overfitting as the number of factors is increased. For $\hat{K}=20$
the variance is also very high, since the four spurious features fit
noise. Using an ARD prior on the features (AFA) improves the
performance, and overfitting no longer occurs. The reconstruction error
is actually less for $\hat{K}=20$, but this is an artifact due to the
reconstruction error not penalizing additional spurious features in the
inferred $\G$. The Sparse PCA (SPCA) of \citet{TibshiraniSPCA2006}
shows improved reconstruction compared to the nonsparse methods (FA and
AFA), but does not perform as well as the Bayesian sparse models.
Sparse factor analysis (SFA), the finite version of the full infinite
model, performs very well. The Bayesian Factor Regression Model (BFRM)
performs significantly better than the ARD factor analysis (AFA), but
not as well as our sparse model (SFA). It is interesting that for BFRM
the reconstruction error decreases significantly with increasing $\hat
{K}$, suggesting that the default priors may actually encourage too
much sparsity for this data set. Fokoue's method (FOK) only performs
marginally better than AFA, suggesting that this ``soft'' sparsity
scheme is not as effective at finding the underlying sparsity in the
data. Overfitting is also seen, with the error increasing with $\hat
{K}$. This could potentially be resolved by placing an appropriate per
factor ARD-like prior over the scale parameters of the Gamma
distributions controlling the precision of elements of $\G$. Finally,
the Nonparametric Sparse Factor Analysis (NSFA) proposed here and in
\citet{daume08ihfrm} performs very well. With fixed $\alpha=1$ (a1) or
inferring $\alpha$, we see very similar performance. Using the soft
coupling (sc) variant which shares power between dimensions when
fitting the noise variances seems to reduce the variance of the
sampler, which is reasonable in this example since the noise was in
fact isotropic.

\begin{figure}

\includegraphics{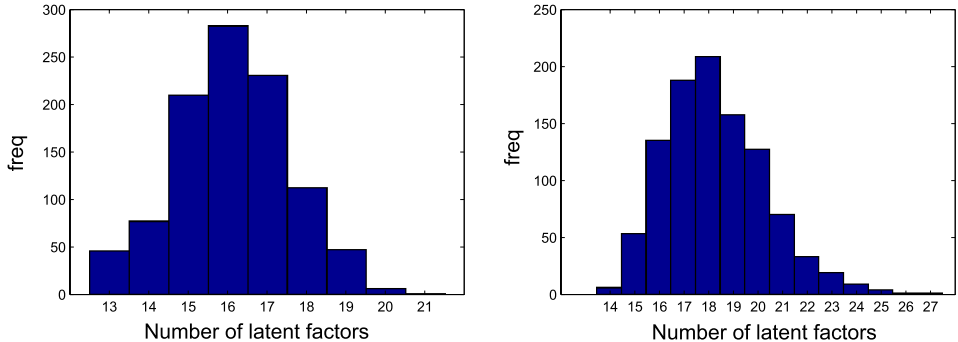}

% syn_test_hist_K.pdf: 691x239 pixel, 72dpi, 24.38x8.43 cm, bb=0 0 691
%239
\caption{Histograms of the number of latent features inferred by the
nonparametric sparse FA sampler for the last 100 samples out of 1000.
\emph{Left}: With $\alpha=1$. \emph{Right}: Inferring $\alpha$.}
\label{fig:syn_test_hist_K}
\end{figure}

Since the reconstruction error does not penalize spurious factors, it
is important to check that NSFA is not scoring well simply by inferring
many additional factors. Histograms for the number of latent features
inferred for the nonparametric sparse model are shown in Figure \ref
{fig:syn_test_hist_K}. This represents an approximate posterior over
$K$. For fixed $\alpha=1$ the distribution is centered around the true
value of $K=16$, with minimal bias ($\mathbb{E}K=16.1$). The variance
is significant (standard deviation of 1.46), but is reasonable
considering the noise level ($\mathrm{SNR}=10$) and that in some of the random
data sets, elements of $\Z$ which are 1 could be masked by very small
corresponding values of $\G$. This hypothesis is supported by the
results of a similar experiment where $\G$ was set equal to $\Z$. In
this case, the sampler always converged to at least 16 features, but
would also sometimes infer spurious features from noise (results not
shown). When inferring $\alpha$ some bias and skew are noticeable. The
mean of the posterior is now at $18.3$ with standard deviation $2.0$,
suggesting there is little to gain from sampling $\alpha$ in this data.

%f5 ###

%s5.2 ###
\subsection{Convergence}

\begin{figure}
\begin{tabular}{@{}c@{\ }c@{\ }c@{}}

\includegraphics{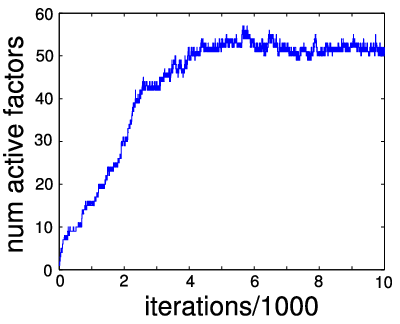}
&\includegraphics{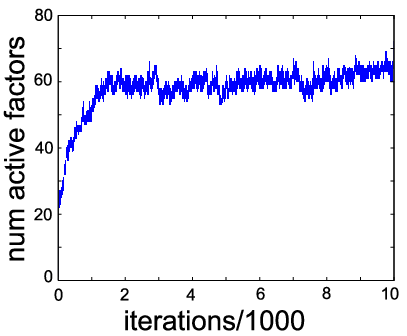}&\includegraphics{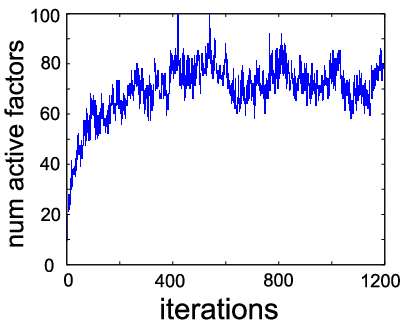}\\
(a)&(b)&(c)
\end{tabular}
\caption{The effect of different proposal distributions for the number
of new features. \textup{(a)}~Prior. \textup{(b)}~Prior plus $0.1\mathbb{I}(\kappa=1)$. \textup{(c)} Factor $\lambda=50$.} \label{fig:new_features_mixing}
\end{figure}

NSFA can suffer from slow convergence if the number of new features is
drawn from the prior. Figure \ref{fig:new_features_mixing} shows how
the different proposals for $\kappa_d$ effect how quickly the sampler
reaches a sensible number of features. If we use the prior as the
proposal distribution, mixing is very slow, taking around 5000
iterations to converge, as shown in Figure \ref{fig:new_features_mixing}(a). If a
mass of $0.1$ is added at $\kappa_d=1$ [see equation (\ref
{eq:kproposal})], then the sampler reaches the equilibrium number of
features in around 1500 iterations, as shown in Figure \ref
{fig:new_features_mixing}(b). However, if we try to add features even faster,
for example, by setting the factor $\lambda=50$ in equation (\ref
{eq:kproposal}), then the sampling noise is greatly increased, as shown
in Figure \ref{fig:new_features_mixing}(c), and the computational cost also
increases significantly because so many spurious features are proposed
only to be rejected.

%f6 ###

%s5.3 ###
\subsection{Biological data: \emph{E. Coli} time-series dataset}

To assess the performance of each algorithm on the biological data
where no ground truth is available, we calculated the test set log
likelihood under the posterior. Ten percent of entries from $\Y$ were
removed at random, ten times, to give ten data sets for inference. We
do not use mean square error as a measure of predictive performance
because of the large variation in the signal to noise ratio across gene
expression level probes.

%f7 ###
\begin{figure}
\begin{tabular}{@{}cc@{}}

\includegraphics{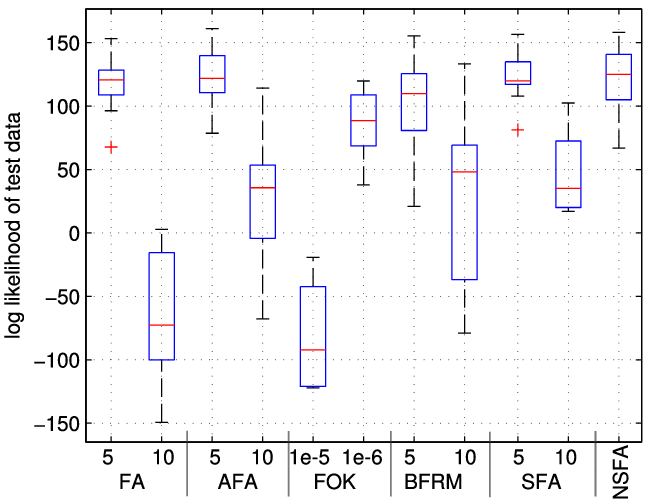}
&\includegraphics{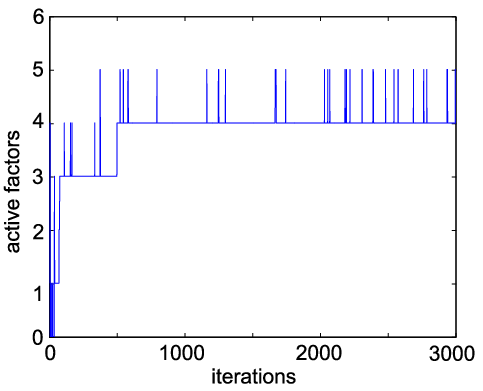}\\
(a)&(b)
\end{tabular}
\caption{Results on \emph{E. Coli} time-series data set from Kao et~al. (\protect\citeyear{Kao2004})
($N=24$, $D=100$, 3000 MCMC iterations). \textup{(a)} Log likelihood of test data under each model based on the
last 100 MCMC samples. The boxplots show variation across 10 different
random splits of the data into training and test
sets. \textup{(b)} Number of active latent features during a typical MCMC run
of the NSFA model.}\label{fig7}
\end{figure}

The test log likelihood achieved by the various algorithms on the \emph
{E. Coli} data set from \citet{Kao2004}, including 100 genes at 24
time-points, is shown in Figure \ref{fig7}(a). On this simple
data set incorporating sparsity doesn't improve predictive performance.
Overfitting the number of latent factors does damage performance,
although using the ARD or sparse prior alleviates the problem. Based on
predictive performance of the finite models, five is a~sensible number
of features for this data set: the NSFA model infers a~median number of
4 features, with some probability of there being 5, as shown in Figure
\ref{fig7}(b).

%f8 ###
\begin{figure}
\begin{tabular}{@{}c@{}}

\includegraphics{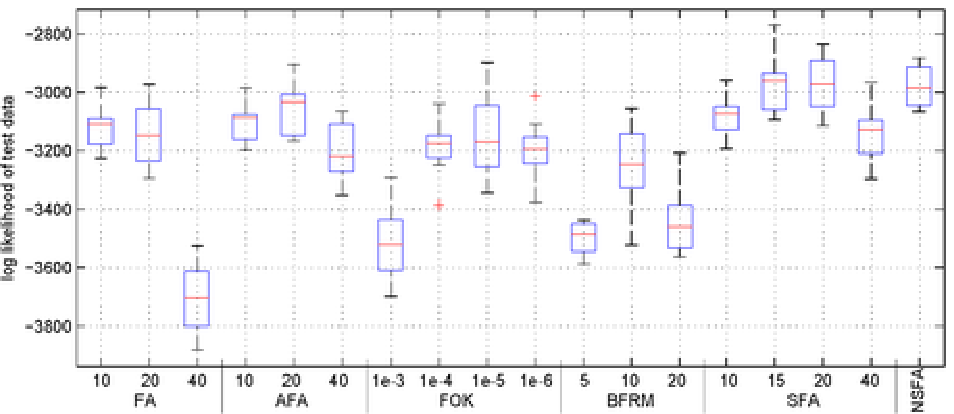}
\\
(a)\\

\includegraphics{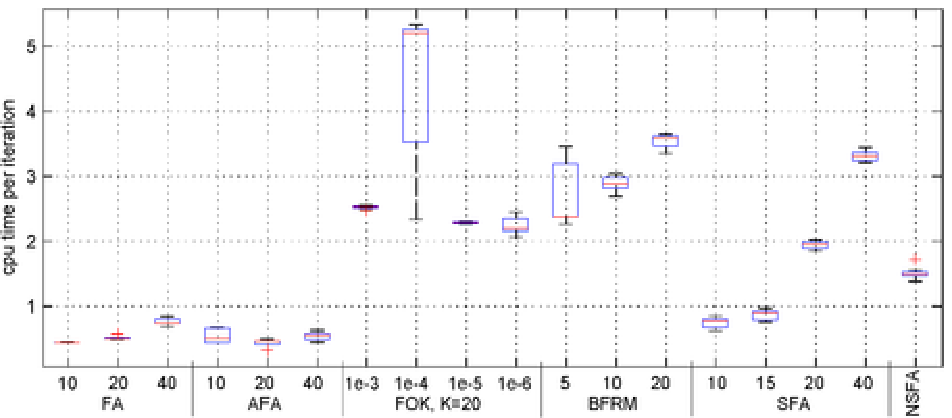}
\\
(b)\\

\includegraphics{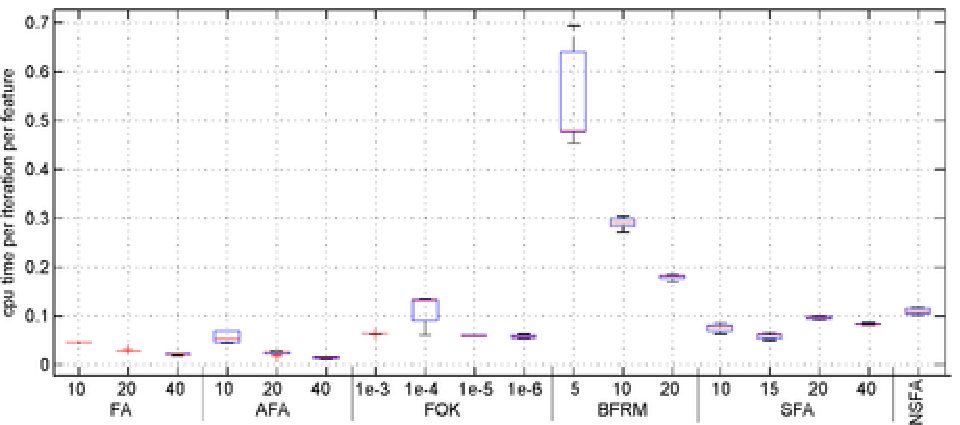}
\\
(c)
\end{tabular}
\vspace*{-4pt}
\caption{Results on breast cancer data set ($N=251$, $D=226$, 3000 MCMC
iterations). \textup{(a)} Predictive performance: log likelihood of test (the 10\%
missing) data under each model based on the last 100 MCMC samples.
Higher values indicate better performance. The boxplots show variation
across 10 different random splits of the data into training and test
sets. \textup{(b)} CPU time (in seconds) per iteration, averaged across the
3000 iteration run. \textup{(c)} CPU time (in seconds) per iteration divided by the number
of features at that iteration, averaged across all
iterations.}\label{fig8}
\vspace*{-6pt}
\end{figure}

%s5.4 ###
\subsection{Breast cancer data set} We assess these algorithms in
terms of \emph{predictive performance} on the breast cancer data set
of \citet{westBFRM2008}, including 226 genes across 251 individuals. We
find that all the finite models are sensitive to the choice of the
number of factors, $K$. The samplers were found to have converged after
around 1000 samples according to standard multiple chain convergence
measures, so 3000 MCMC iterations were used for all models. The
predictive log likelihood was calculated using the final 100 MCMC
samples. Figure \ref{fig8}(a) shows test set log likelihoods for 10
random divisions of the data into training and test sets. Factor
analysis (FA) shows significant overfitting as the number of latent
features is increased from 20 to 40. Using the ARD prior prevents this
overfitting (AFA), giving improved performance when using 20 features
and only slightly reduced performance when 40 features are used. The
sparse finite model (SFA) shows an advantage over AFA in terms of
predictive performance as long as underfitting does not occur:
performance is comparable when using only 10 features. However, the
performance of SFA is sensitive to the choice of the number of factors,
$K$. The performance of the sparse nonparametric model (NSFA) is
comparable to the sparse finite model when an appropriate number of
features is chosen, but avoids the time consuming model selection
process. Fokoue's method (FOK) was run with $K=20$ and various settings
of the hyperparameter $d$ which controls the overall sparsity of the
solution. The model's predictive performance depends strongly on the
setting of this parameter, with results approaching the performance of
the sparse models (SFA and NSFA) for $d=10^{-4}$. The performance of
BFRM on this data set is noticeably worse than the other sparse models.

We now consider the \emph{computation cost} of the algorithms. As
described in Section \ref{sec:inference}, sampling $\Z$ and $\G$
takes order $O(NKD)$ operations per iteration, and sampling $\X$ takes
$O(K^2+K^3+ND)$. However, for the moderate values encountered for data
sets 1 and 2, the main computational cost is sampling the nonzero
elements of $\G$, which takes $O((1-s)DK)$ where $s$ is the sparsity
of the model. Figure \ref{fig8}(c) shows the mean CPU time per
iteration divided by the number of features at that iteration.
Naturally, straight FA is the fastest, taking only around $0.025s$ per
iteration per feature. The value increases slightly with increasing
$K$, suggesting that here the $O(K^2 D+K^3)$ calculation and inversion
of $\bolds{\lambda}$, the precision of the conditional on $\X$, must be
contributing. The computational cost of adding the ARD prior is
negligible (AFA). The CPU time per iteration is just over double for
the sparse finite model (SFA), but the cost actually decreases with
increasing $K$, because the sparsity of the solution increases to avoid
overfitting. There are fewer nonzero elements of $\G$ to sample per
feature, so the CPU time \emph{per feature} decreases. The CPU time
per iteration per feature for the nonparametric sparse model (NSFA) is
somewhat higher than for the finite model because of the cost of the
feature birth and death process. However, Figure \ref{fig8}(b)
shows the absolute CPU time per iteration, where we see that the
nonparametric model is only marginally more expensive than the finite
model of appropriate size $\hat{K}=15$ and cheaper than choosing an
unnecessarily large finite model (SFA with $K=20$,  40). Fokoue's method
(FOK) has comparable computational performance to the sparse finite
model, but interestingly has increased cost for the optimal setting of
$d=10^{-4}$. The parameter space for FOK is continuous, making search
easier but requiring a normal random variable for every element of $\G
$. BFRM pays a considerable computational cost for both the
hierarchical sparsity prior and the DP prior on $\X$. SPCA was not run
on this data set, but results on the synthetic data in Section \ref
{sec:syndata} suggest it is somewhat faster than the sampling methods,
but not hugely so. The computational cost of SPCA is $ND^2 +
mO(D^2K+DK^2+D^3)$ in the $N>D$ case (where $m$ is the number of
iterations to convergence) and $ND^2+mO(D^2K+DK^2)$ in the $D>N$ case,
taking the limit $\lambda\rightarrow\infty$. In either case an
individual iteration of SPCA is more expensive than one sampling
iteration of NSFA (since $K<D$), but fewer iterations will generally be
required to reach convergence of SPCA than are required to ensure
mixing of NSFA.

%f9 ###
\begin{figure}

\includegraphics{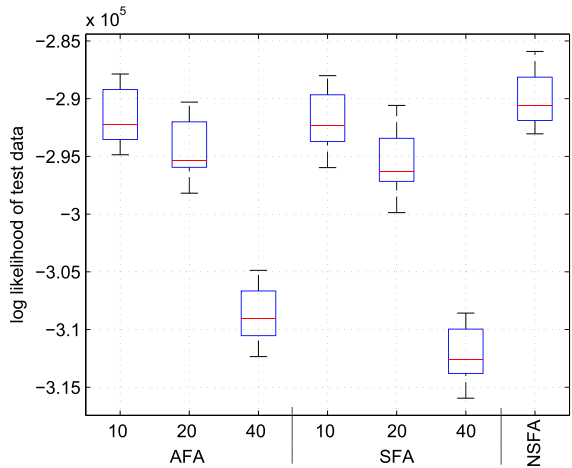}

\caption{Test set log likelihoods on Prostate cancer data set from
Yu et~al. (\protect\citeyear{Yu2004}), including 12557 genes across 171 individuals (1000 MCMC
iterations).} \label{fig:big_ll}
\end{figure}

%s5.5 ###
\subsection{Prostate cancer data set} Figure \ref{fig:big_ll} shows
the predictive performance of AFA, FOK and NSFA on the prostate cancer
data set of \citet{Yu2004}, for ten random splits into training and
test data. The boxplots show variation from ten random splits into
training and test data. The large number of genes ($D=12557$ across
$N=171$ individuals) in this data set makes inference slower, but the
problem is manageable since the computational complexity is linear in
the number of genes. Despite the large number of genes, the appropriate
number of latent factors, in terms of maximizing predictive
performance, is still small, around 10 (NSFA infers a median of 12
factors). This may seem small relative to the number of genes, but it
should be noted that the genes included in the breast cancer and \emph
{E. Coli} data sets are those capturing the most variability.
Surprisingly, SFA actually performed slightly worse on this data set
than AFA. Both are highly sensitive to the number of latent factors
chosen. NSFA, however, gives better predictive log likelihoods than
either finite model for any fixed number of latent factors $K$. Running
1000 iterations of NSFA on this data set takes under 8 hours. BFRM and
FOK were impractically slow to run on this data set.

%s6 ###
\section{Discussion}

We have seen that in both the \textit{E. Coli} and breast cancer data
sets that sparsity can improve predictive performance, as well as
providing a more easily interpretable solution. Using the IBP to
provide sparsity is straightforward, and allows the number of latent
factors to be inferred within a well-defined theoretical framework.
This has several advantages over manually choosing the number of latent
factors. Choosing too few latent factors damages predictive
performance, as seen for the breast cancer data set. Although choosing
too many latent factors can be compensated for by using appropriate
ARD-like priors, we find this is typically more computationally
expensive than the birth and death process of the IBP. Manual model
selection is an alternative but is time consuming. Finally, we show
that running NSFA on full gene expression data sets with 10000$+$ genes
is feasible so long as the number of latent factors remains relatively
small. An interesting direction for this research is how to incorporate
prior knowledge, for example, if certain transcription factors are
known to regulate specific genes. Incorporating this knowledge could
both improve the performance of the model and improve interpretability
by associating latent variables with specific transcription factors.
Another possibility is incorporating correlations in the Indian Buffet
Process, which has been proposed for simpler models [\citet
{finale-corribp}; \citet{Courvilleaninfinite}]. This would be appropriate in a
gene expression setting where multiple transcription factors might be
expected to share sets of regulated genes due to common motifs.
Unfortunately, performing MCMC in all but the simplest of these models
suffers from slow mixing.

\section*{Acknowledgments}

We would like to thank the anonymous reviewers for helpful comments.

\begin{supplement}%[id=suppA]
% \sname{Supplement A}
\stitle{Graphs of precision and recall for the synthetic data experiment.}
\slink[doi]{10.1214/10-AOAS435SUPP}
\slink[url]{http://lib.stat.cmu.edu/aoas/435/supplement.pdf}
\sdatatype{.pdf}
\sdescription{The precision and recall of active elements of the $\Z$
matrix achieved by each algorithm (after thresholding for the nonsparse
algorithms) on the synthetic data experiment, described in
Section \ref{sec:syndata}. The results are consistent with the
reconstruction error.}
\end{supplement}

%suskaldyti doi

\printaddresses

\end{document}